\documentclass{article}
\usepackage{spconf,amsmath,amssymb,graphicx,hyperref}

\ninept

\usepackage{braket}

\usepackage{tikz}
\usetikzlibrary{quantikz2}
\usetikzlibrary{shapes,arrows,chains}

\usepackage{orcidlink}

\newcommand \R {\mathbb{R}}
\newcommand \C {\mathbb{C}}

\newcommand \Delay {\mathbf{\Lambda}}

\newcommand \mbf[1] {\mathbf{#1}}

\title{STATE-SPACE-BASED FIR FILTERING ON A QUANTUM COMPUTER}
\name{Roope Salmi$^{1}$\orcidlink{0009-0007-3579-0726} \quad Davide Rocchesso$^{2}$\orcidlink{0000-0002-0849-7766} \quad Vesa Välimäki$^{1}$\orcidlink{0000-0002-7869-292X}}
\address{$^{1}$ Acoustics Lab, Dept. of Information and Communications Eng., Aalto University, Espoo, Finland \\ 
$^{2}$ Università degli studi di Milano Statale, Milan, Italy}

\begin{document}
\maketitle
\begin{abstract}
Many signal processing tasks require intensive computations. Quantum computing promises to accelerate certain tasks, but algorithms must be designed around the limitations of quantum mechanics. This paper provides a quantum implementation of finite impulse response (FIR) filters, which are a widely used tool in classical signal processing. The filter can be parallelized and composed as part of larger quantum algorithms, with potential for speedups using quantum amplitude estimation and future fault-tolerant quantum hardware. To accomplish these properties, we introduce a state-space framework wherein sample-based signal processing is performed with unitary quantum circuits. We present the quantum delay gate as the quantum analog of the delay line in discrete-time signal processing. Unitary circuits intrinsically describe lossless systems, but we also implement lowpass and other bounded filters by emulating a projection operator. The FIR filter implementation is tested and demonstrated with a quantum circuit simulator.
\end{abstract}
\section{Introduction}
\label{sec:intro}

Quantum computing has notable applications in fields such as number theory~\cite{shor_algorithms_1994}, optimization~\cite{moll2018}, and chemistry~\cite{mcardle_quantum_2020}. Many quantum algorithms have been found that promise to be more efficient than known classical approaches. Some algorithms target current noisy intermediate-scale quantum (NISQ) hardware~\cite{preskill_quantum_2018}, but most require larger, fault-tolerant quantum computers, which remain under development~\cite{awschalom_challenges_2025}.
Applications of quantum computing to signal processing are being explored, for example, in quantum image processing (QIMP)~\cite{yan_survey_2016}. For one-dimensional (1D) signals such as audio~\cite{wang2016}, no definite quantum speedups have been claimed.
This paper advances efficient quantum algorithms for 1D signal processing by enabling longer computations.

Quantum signal processing (QSP) has conflicting definitions in the literature. It refers both to the general study of signal processing on or inspired by quantum systems~\cite{eldar2002, yin_quantum_2021, shukla_quantum_2023, nair_short-time_2025}, and to a particular algorithm that uses signal processing principles~\cite{low2017}. Here, QSP refers to the former, i.e., signal processing on a quantum computer.

Taking inspiration from QIMP, various strategies for representing a 1D signal as a quantum state have been proposed~\cite{wang2016, yan_flexible_2018, itaborai_quantum_2022}. Since the dimension of a quantum state vector (QSV) grows exponentially with the number of qubits~\cite{nielsen_quantum_2010}, long signals can be represented compactly. Certain operations can be applied to multiple samples in parallel. However, signal preparation, processing, and readout are constrained by the postulates of quantum mechanics~\cite{itaborai_quantum_2022}.
Despite these constraints, block-based processing has been proposed with the quantum Fourier transform~\cite{yin_quantum_2021, nair_short-time_2025, sharma_signal_2023, papageorgiou_2026}, with the Walsh--Hadamard transform~\cite{shukla_quantum_2023}, and to implement delay~\cite{aguado_2026}. Delays and memory effects also emerge in measurement processes of open quantum systems~\cite{evgi_2026}.
Sample-based time-domain processing has been studied in the form of quantum feedback delay networks~\cite{rocchesso2025} and finite impulse response (FIR) filters~\cite{majumdar2025, tseng_finite_2026}.

The existing time-domain QSP methods are limited in how they can be composed. 
In the quantum FIR filter implementation by Majumdar et al.~\cite{majumdar2025}, multiple input samples are consumed to produce one output sample, and the input samples cannot be reused. The filter therefore cannot be connected to other signal processing components without significant loss of gain or converting to a classical representation in between, which is expensive and prone to noise accumulation~\cite{itaborai_quantum_2022}.
Quantum speedups arising from parallel operations are only achievable if the number of useful arithmetic operations performed is much greater than the number of converted samples.

This paper develops a framework for composable time-domain signal processing in a closed quantum system. Any lossless linear system can be realized as a unitary quantum operator. Additionally, dissipative linear systems can be emulated by employing a sufficiently large state space. Since the total energy in the quantum state cannot be increased, it is important to ensure that there are no unnecessary gain losses in the designed systems. We introduce concrete implementations of delay lines and FIR filters as quantum circuits. The FIR filter implementation is based on a unitary decomposition for lossless FIR filterbanks by Vaidyanathan~\cite{vaidyanathan_passive_1986,vaidyanathan_multirate_1993}. It supports all bounded FIR filters, but a power complementary filter must be known. Quantum amplitude estimation (QAE)~\cite{brassard_quantum_2000} may be used to read output samples efficiently. We expect that parallel variants of this filter could be used to speed up computation of large, interconnected linear systems.

The remainder of this paper is structured as follows. Sec.~\ref{sec:quantum-basics} is a brief introduction to quantum computing and notation, Sec.~\ref{sec:representation} presents the framework for composable QSP, Sec.~\ref{sec:fir-filters} implements FIR filters in this framework, Sec.~\ref{sec:simulation} tests our filter implementation in a simulator, and Sec.~\ref{sec:conclusion} concludes.

\section{Quantum States and Circuits}
\label{sec:quantum-basics}

The qubit is an abstraction of a two-level quantum system, whose state is written in bra-ket notation as $\ket{\psi} = a\ket{0} + b\ket{1}$, where $\ket{0}$ and $\ket{1}$ are basis states and $a, b \in \C$ are probability amplitudes~\cite{nielsen_quantum_2010}. The amplitudes correspond to the probabilities $P(0) = |a|^2, P(1) = |b|^2$ of measuring $\ket{0}$ or $\ket{1}$, respectively, if the qubit is observed. A quantum system of $q$ qubits is understood as the tensor product of $q$ two-dimensional Hilbert spaces. Each of the $2^q$ binary combinations is associated with a distinct probability amplitude. A QSV is always a unit vector, so that the measurement probabilities sum to $1$. Instead of binary notation, we write $\ket{x}$, where $0 \le x < 2^q$ to describe the basis states of a $q$-qubit register.

Quantum states can be manipulated using gates or sequences of gates, called circuits. Besides measurement, all runnable gates are described by unitary matrices. We make use of the single-qubit gates
\begin{equation}
    \mbf{H} =
    \frac{1}{\sqrt{2}}\begin{bmatrix}
        1 & 1 \\
        1 & -1
    \end{bmatrix},
    \quad
    \mbf{R}_y(\theta) =
    \begin{bmatrix}
        \cos(\theta/2) & \sin(\theta/2) \\
        -\sin(\theta/2) & \cos(\theta/2)
    \end{bmatrix}.
\end{equation}
The Hadamard gate $\mbf{H}$ generates the uniform superposition $\ket{+} = \mbf{H}\ket{0} = (\ket{0} + \ket{1}) / \sqrt{2}$. The gate $\mbf{R}_y(\theta)$ is a rotation around the $y$-axis of the Bloch sphere~\cite{nielsen_quantum_2010} by the angle $\theta$. Note that $\theta$ is twice that of the angle in a conventional planar rotation matrix.
The multi-qubit transposition gate $\mbf{T} = \mbf{I} - \ket{x}\bra{x} - \ket{y}\bra{y} + \ket{y}\bra{x} + \ket{x}\bra{y}$, where $\mbf{I}$ is the identity matrix, exchanges two amplitudes $\ket{x}$ and $\ket{y}$ without affecting any other amplitudes. It can be implemented using $\mathcal{O}(q)$ elementary gates, where $q$ is the total number of qubits~\cite{fuchs_compact_2025}.

Given a state prepared by a unitary $\mbf{U}$, the Hadamard test~\cite{aharonov_polynomial_2006} estimates the real or imaginary part of a probability amplitude. For the real part, we prepare an ancillary qubit in superposition $\ket{+}$, apply the controlled unitary $\mbf{U}$, and measure the ancilla qubit in the computational basis after a Hadamard gate. While direct sampling requires $\mathcal{O}(1/\varepsilon^2)$ measurements to estimate $P(0)$ within error $\varepsilon$, QAE~\cite{brassard_quantum_2000} provides a quadratic speedup for sample readout. 

\section{Representation of Signals and Systems}
\label{sec:representation}

In this work, signals are represented in the probability amplitudes of a quantum state. Manipulations of the signal are restricted to unitary transformations, but various linear signal processing techniques are shown to be possible.

\subsection{Signal representation as a quantum state}

Let $x(n)$ be a real or complex-valued discrete-time signal with length $N$. We assume it is normalized to $\lVert x(n) \rVert^2=1$ unless stated otherwise. The signal is represented as a quantum state by
\begin{equation}
    \label{eq:qpam}
    \ket{A_x} = \sum_{n=0}^{N-1} x(n) \ket{n}.
\end{equation}
At least $\lceil \log_2(N)\rceil$ qubits are needed for the sample position $\ket{n}$. This representation is similar to quantum probability amplitude modulation (QPAM)~\cite{itaborai_quantum_2022}, although we do not require the values to be positive and real. Indeed, negative and complex-valued samples are supported, since the Hadamard test makes them observable.

A signal can be prepared as a quantum state using general-purpose state preparation routines~\cite{itaborai_quantum_2022}. They require $\mathcal{O}(N)$ elementary gates to prepare an arbitrary QSV with $N$ elements~\cite{sun_asymptotically_2023}. Ready implementations are available in libraries such as Qiskit~\cite{javadi_2024}. Sample values can be read from the state via direct sampling or QAE of the Hadamard test.

\subsection{State-space realization of linear systems}

Consider a discrete-time multi-input, multi-output linear system with $L$ inputs and outputs and $S$ system state variables. The input and output at sample $n$ are vectors $\mbf{x}(n), \mbf{y}(n) \in \C^{L}$. The system state vector (SSV), $\mbf{s}(n) \in \C^{S}$, is stored between samples.
The system is described by the difference equations
\begin{equation}
\begin{aligned}
    \mbf{y}(n) &= \mbf{A}\,\mbf{x}(n) + \mbf{B}\,\mbf{s}(n) \\
    \mbf{s}(n+1) &= \mbf{C}\,\mbf{x}(n) + \mbf{D}\,\mbf{s}(n),
\end{aligned}
\end{equation}
with the matrices $\mbf{A} \in \C^{L \times L}, \mbf{B} \in \C^{L \times S}, \mbf{C} \in \C^{S \times L},$ and $\mbf{D} \in \C^{S \times S}$. The realization matrix~\cite{vaidyanathan_multirate_1993} of the system is
\begin{equation}
    \label{eq:realization-matrix}
    \mbf{U} = 
    \begin{bmatrix}
        \mbf{A} & \mbf{B} \\
        \mbf{C} & \mbf{D}
    \end{bmatrix}.
\end{equation}
Contrary to convention~\cite{vaidyanathan_multirate_1993, schlecht_allpass_2021}, we order the input-output (IO) vector before the SSV. This makes it easier to identify the IO bases in a quantum state.

If $\mbf{U}$ is unitary, the system can be implemented as a quantum circuit~\cite{nielsen_quantum_2010} on $\lceil \log_2(L + S) \rceil$ qubits. Applying the operator $\mbf{U}$ to the state $\ket{\mbf{x}(n)} + \ket{\mbf{s}(n)}$ yields $\ket{\mbf{y}(n)} + \ket{\mbf{s}(n+1)}$, where $\ket{\mbf{x}(n)}$ and $\ket{\mbf{y}(n)}$ belong to the same subspace, while $\ket{\mbf{s}(n)}$ belongs to an orthogonal subspace. If we can route the output $\ket{\mbf{y}(n)}$ to another subspace and replace it with the next input $\ket{\mbf{x}(n+1)}$, we can process the next sample by applying $\mbf{U}$ again, and so on.

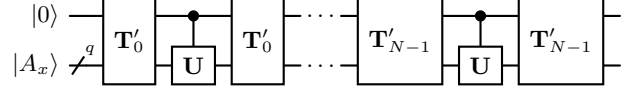
\begin{figure}
    \centering
    \begin{quantikz}[row sep=0.5em, column sep=0.7em]
        \lstick{$\ket{0}$} & & \gate[2, disable auto height]{\mbf{T}'_{0}} & \ctrl{1} & \gate[2, disable auto height]{\mbf{T}'_{0}} & \ \ldots\  & \gate[2, disable auto height]{\mbf{T}'_{N-1}} & \ctrl{1} & \gate[2, disable auto height]{\mbf{T}'_{N-1}} & \\
        \lstick{$\ket{A_x}$} & \qwbundle{q} & & \gate{\mbf{U}} & & \ \ldots\ & & \gate{\mbf{U}} & & \\
    \end{quantikz}
    \caption{A circuit to process all samples $n=0, \dots, N-1$ of a signal encoded as $\ket{A_x}$ using the SISO realization matrix $\mbf{U}$. The bottom register consists of $q=\lceil \log_2(\max(N, S+1)) \rceil$ qubits. }
    \label{fig:fullsigproc}
\end{figure}

As an example, Fig.~\ref{fig:fullsigproc} shows how to process a signal encoded as in \eqref{eq:qpam} using a single input, single output (SISO) system, $L=1$, with realization matrix $\mbf{U}$. We use an additional qubit to distinguish between the signal buffer and the SSV, which occupy orthogonal subspaces in a $N+L+S=N+S+1$-dimensional Hilbert space.  The initial state is $\ket{0} \ket{A_x}$. The following procedure processes sample $n$. First, the sample $x(n)$ is moved to the IO basis with a transposition gate $\mbf{T}'_n$ between $\ket{0}\ket{n}$ and $\ket{1}\ket{0}$. Then, we apply the realization matrix $\mbf{U}$ controlled on the first qubit. The resulting sample $y(n)$ is moved back to the signal buffer with $\mbf{T}'_n$.

% For sample processing:
% T'_n : |0>|n> <-> |1>|0>
% Later, for delay gates:
% T_k : |0>|1> <-> |k>|0>

The whole signal is processed in Fig.~\ref{fig:fullsigproc} by repeating the above procedure for all samples $n=0, 1, \dots, N-1$. The final state is $\ket{0} \ket{A_y} + \ket{1}\ket{\mbf{s}(N)}$, where $\ket{A_y}$ represents the encoding of the output signal $y(n)$. The residual SSV $\ket{\mbf{s}(N)}$ may be nonzero, and therefore, some energy may have been removed from the signal $x(n)$ to produce the signal $y(n)$. This fact enables the implementation of a wide variety of filters. Instead of directly processing a signal, the system can also be composed or connected in feedback with other processing using appropriate transposition gates.

\subsection{Parallel processing}
\label{ssec:parallel-processing}

Quantum computers naturally operate on a superposition in parallel~\cite{nielsen_quantum_2010}. In this way, multiple instances of a linear system can be evaluated simultaneously for the same cost as evaluating the system once.
Let $\mbf{x}_j(n), \mbf{y}_j(n), \mbf{s}_j(n)$ be, respectively, the IO and SSV of the system instance $j=0, \dots, 2^r-1$.
With $r$ additional qubits, we may prepare the superposition
$
\sum_{j} \ket{j} \otimes \left[\ket{\mbf{x}_j(n)} + \ket{\mbf{s}_j(n)}\right]
$.
For this to be a valid quantum state, total energy across all instances must sum to one.
The operator $\mbf{I} \otimes \mbf{U}$, i.e., the realization matrix applied to the system IO and state qubits, transforms the state into
$
\sum_{j} \ket{j} \otimes \left[\ket{\mbf{y}_j(n)} + \ket{\mbf{s}_j(n+1)}\right]
$.
Thus, the sample $n$ is processed for all $2^r$ instances in parallel.

\subsection{Delay gate}

A SISO delay of $d$ samples is achieved in the state space form with $S=d$ state variables and a realization matrix $\mbf{U}$ that is a cyclic permutation. The delay is applied in the time domain, rather than to a whole signal block, as done by Aguado-Y{\'a}{\~n}ez et al.~\cite{aguado_2026}. In filters, we wish to compose multiple delay lines in the same system. This is organized using a buffer register of $b$ qubits to distinguish the work buffer ($k=0$) and the state spaces of each distinct delay line ($k > 0$). A further $q=\lceil \log_2(d) \rceil$ qubits are needed for the delay buffer, but these can be shared among other buffers in superposition.

\begin{figure}
    \centering
    \begin{quantikz}[column sep={0.6em}]
        & \gate{\Delay^d} &
    \end{quantikz}
    $\equiv$
    \begin{quantikz}[align equals at=1.5, row sep = 0.5em]
        & \qwbundle{b} &[-0.8em] \gate[2, disable auto height]{\mbf{T}_{k}} & \measure{k} \wire[d]{q} &[-0.8em] \\
        & \qwbundle{q} & & \gate{+1 \bmod d} & \\
    \end{quantikz}
    \caption{A $d$-sample delay gate in the $z$-domain, and its realization in buffer $k$ with a tranposition $\mbf{T}_k$ and a controlled increment gate. The circled $k$ indicates multi-qubit control on the buffer register.}
    \label{fig:delaygate}
\end{figure}
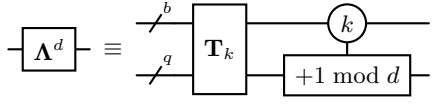

We define the $d$-sample delay gate in the $z$-domain as
$\mbf{\Lambda}^d(z) = \operatorname{diag}(1, z^{-d})$. Similar notation is used by Vaidyanathan~\cite{vaidyanathan_multirate_1993}. The amplitude $\ket{1}$ is multiplied by $z^{-d}$, while $\ket{0}$ is left intact. The right side of Fig.~\ref{fig:delaygate} shows a possible realization. We assume only one qubit is used in the work buffer. The transposition gate $\mbf{T}_k$ exchanges the amplitudes $\ket{0}\ket{1} \leftrightarrow \ket{k}\ket{0}$ between the work and delay buffers. The delay buffer is then permuted with an increment gate. It is applied controlled on the buffer register state $\ket{k}$, so that it maps $\ket{k}\ket{x} \mapsto \ket{k}\ket{n+1 \bmod d}$ for all $0 \le n < d$. The increment gate can be implemented efficiently with $\mathcal{O}(\log d)$ gates~\cite{vandaele_asymptotically_2026}. For a unit delay $\Delay(z)$, the increment gate is omitted.
After $d$ applications, an input sample returns to $\ket{k}\ket{0}$, and is moved back to the work buffer by the transposition gate on the $d+1$th application.

\subsection{Projection gate}

When the realization matrix \eqref{eq:realization-matrix} is unitary, the resulting system is lossless~\cite{vaidyanathan_passive_1986}. In order to implement bounded filters, i.e., SISO systems whose frequency response is such that $|H(e^{i\omega})| \le 1$, where $i$ is the imaginary unit and $\omega$ is the angular frequency, we extend the realization matrix to contractions. An operator $\mbf{V}$ is a contraction if $\lVert \mbf{V}\ket{\psi} \rVert \le \lVert \ket{\psi} \rVert$ for all states $\ket{\psi}$. Contractions cannot be directly implemented as quantum circuits, but we use an approach similar to the Sz.-Nagy dilation theorem~\cite{schaffer_unitary_1955} to emulate them.

The single-qubit projection operator $\ket{0}\bra{0}$, which discards the amplitude~$\ket{1}$, is a contraction. It can be emulated with a long delay gate that moves the amplitude $\ket{1}$ away:
\begin{equation}
    \ket{0}\bra{0} \simeq \Delay^p(z).
\end{equation}
The integer $p$ is an upper bound for the number of samples the system can process: If more than $p$ samples were processed, nonzero samples would be read back from the delay. Choosing $p$ as a power of two simplifies the realization of the delay gate. The emulation is efficient because only a logarithmic number of qubits and gates are needed relative to $p$.

\section{FIR Filters as Quantum Circuits}
\label{sec:fir-filters}

Using the state-space framework and gates defined in Sec.~\ref{sec:representation}, we can implement FIR filters as quantum circuits. We show a first-order example, which is then generalized to arbitrary orders with a classic cascaded lattice structure by Vaidyanathan~\cite{vaidyanathan_passive_1986}.

\subsection{First-order FIR filter}

Consider the first-order lowpass FIR filter $H(z) = 1/2 + 1/2\,z^{-1}$. It can be implemented as a quantum circuit using two Hadamard gates, a unit delay gate, and a projection gate. On a single qubit in the $z$-domain, let $\mbf{F}(z) = \ket{0}\bra{0} \mbf{H} \Delay(z) \mbf{H}$. IO is performed via the amplitude $\ket{0}$, so the transfer function is
\begin{equation}
    \begin{aligned}
    H(z) &= \bra{0} \mbf{F}(z) \ket{0} = \bra{0} \mbf{H} \Delay(z) \ket{+} \\
    &= \bra{0}\mbf{H} \left(\ket{0} + z^{-1}\ket{1}\right) / \sqrt{2}
    = 1/2 + 1/2\, z^{-1}.
    \end{aligned}
\end{equation}
Without the projection gate $\ket{0}\bra{0}$, the amplitude $\ket{1}$ could remain nonzero and cause errors when processing the subsequent sample.
Replacing the unit delay with $\Delay^d(z)$ results in a comb filter.

The $z$-domain description of the circuit $\mbf{F}(z)$ is realized by replacing the delay and projection gates with the constructions defined earlier. This is detailed below for the general case.

\subsection{General FIR filter}

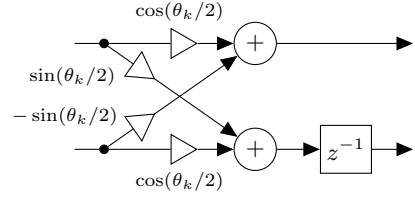
\begin{figure}
    \centering
    \tikzset{%
      dsp/.style = {node distance=1.5cm, >=triangle 45, label distance=0.5mm},
      dblock/.style    = {draw, rectangle, minimum height = 2em,
        minimum width = 2em, inner sep=0.8mm},
      sum/.style      = {draw, circle, inner sep=0.8mm},
      gain/.style     = {draw, regular polygon, regular polygon sides=3, shape border rotate=-90, inner sep=0.8mm, fill=white},
      split/.style = {circle, draw, fill, inner sep=0.4mm}
    }
    \begin{tikzpicture}[dsp, on grid, node distance=1cm]
		\begin{scope}[start chain]
            \node[on chain] (in0a) {};
			\node[node distance=0.5cm, on chain, split] (sp1al) {};
			\node[on chain, gain] (g1aa) [label=above:{\scriptsize $\cos(\theta_k/2)$}] {};
			\node[on chain, sum] (sp1ar) {$+$};

            \node[node distance=1.2cm, on chain] (d2a) {};
            \node[node distance=1cm, on chain] (out2a) {};
		\end{scope}

		\begin{scope}[start chain]
            \node[on chain, below=1.4cm of in0a] (in0b) {};
			\node[node distance=0.5cm, on chain, split] (sp1bl) {};
			\node[on chain, gain] (g1bb) [label=below:{\scriptsize $\cos(\theta_k/2)$}] {};
			\node[on chain, sum] (sp1br) {$+$};

            \node[node distance=1.2cm, on chain, dblock] (d2b) {$z^{-1}$};
            \node[node distance=1cm, on chain] (out2b) {};
		\end{scope}

        \draw[->] (in0a) -- (sp1al) -- (g1aa) -- (sp1ar);
        \draw[->] (in0b) -- (sp1bl) -- (g1bb) -- (sp1br);

		\draw[->] (sp1al) -- (sp1br)
		node [pos=0.25, gain, sloped] {}
		node [pos=0.25, xshift=-9mm, yshift=-1mm] {\scriptsize $\sin(\theta_k/2)$};
		\draw[->] (sp1bl) -- (sp1ar)
		node [pos=0.25, gain, sloped] {}
		node [pos=0.25, xshift=-10mm, yshift=1mm] {\scriptsize $-\sin(\theta_k/2)$};

        \draw[->] (sp1ar) -- (out2a);
        \draw[->] (sp1br) -- (d2b);
        \draw[->] (d2b) -- (out2b);

	\end{tikzpicture}
    \caption{Block diagram of the rotation $\mbf{R}_k = \mbf{R}_y(\theta_k)$ and delay $\Delay(z)$ lattice sections.}
    \label{fig:lattice-sections}
\end{figure}

The cascaded lattice structure by Vaidyanathan~\cite{vaidyanathan_passive_1986} consists of lossless rotation and delay sections, shown in Fig.~\ref{fig:lattice-sections}, which translate directly into unitary quantum gates. Any bounded causal FIR filter can be implemented with this structure without loss of gain. The lattice has two inputs and outputs ($L=2$), of which the second input is unused. The first output gives the desired filter response, while the second output gives a power complementary response, which we discard using a projection gate. The filter is parameterized by a collection of angles, necessitating conversion from direct form coefficients. The presented filter is restricted to real transfer functions, but a complex extension is possible using generalized rotation gates.

\begin{figure}
	\centering

    % Custom command to draw a curly brace above/below columns #1 to #2
    \newcommand{\xgroup}[2]{\draw[decorate,decoration={mirror, brace}] ($(\tikzcdmatrixname-1-#1.west |- \tikzcdmatrixname-row1.south)$) -- ($(\tikzcdmatrixname-1-#2.east |- \tikzcdmatrixname-row1.south)$)}
    \begin{quantikz}[name=cell, column sep=0.7em, execute at end picture={
        \xgroup{2}{3};
        \xgroup{7}{8};
        \xgroup{9}{9};
    }]
		& \gate{\mbf{R}_0} & \gate{\Delay} &
        \gate{\mbf{R}_1} & \gate{\Delay} &
        \ \ldots\  &[0.1em] \gate{\Delay} & \gate{\mbf{R}_M} & \gate{\ket{0}\bra{0}} &
	\end{quantikz} \\[-4pt]
    (a) \\[6pt]
    \renewcommand{\xgroup}[2]{\draw[decorate,decoration={brace}] ($(\tikzcdmatrixname-1-#1.west |- \tikzcdmatrixname-row1.north)$) -- ($(\tikzcdmatrixname-1-#2.east |- \tikzcdmatrixname-row1.north)$)}
    \begin{quantikz}[font=\small, column sep=0.7em, row sep=0.5em, execute at end picture={
        \xgroup{3}{4};
        \xgroup{6}{7};
        \xgroup{8}{9};
    }]
        & \qwbundle{b} &[-0.7em] \measure{0} \wire[d][3]{q} & \gate[4]{\mbf{T}_1} & \ \ldots\  &[0.1em] \gate[4]{\mbf{T}_M} & \measure{0} \wire[d][3]{q} & \gate[4]{\mbf{T}_{M+1}} & \measure{M{+}1} \wire[d]{q} &[-0.7em] \\
        & & & & \ \ldots\  & & & & \gate[3]{+1} & \\[-1.1em]
        \setwiretype{n} \quad\vdots & & & & & & & & & \vdots\quad \\[-0.6em]
        & & \gate{\mbf{R}_0} & & \ \ldots\  & & \gate{\mbf{R}_M} & & &
    \end{quantikz} \\
    (b)

	\caption{\label{fig:fir-circuit} (a) The FIR filter circuit in the $z$-domain and (b) its realization with the delay and projection gates expanded. Braces indicate corresponding groups of gates between the two circuits. Rotations $\mbf{R_k}$ are applied in (b) to the least significant qubit at the bottom.}
\end{figure}
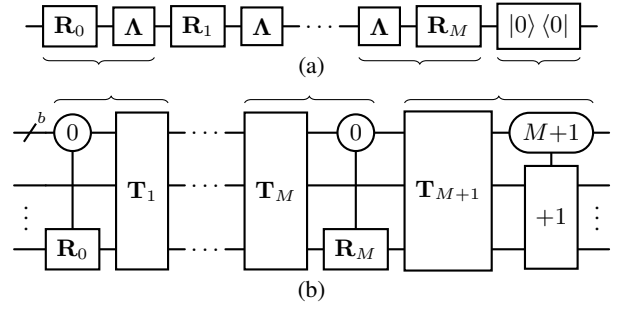

An order $M$ lattice is defined by the transfer matrix~\cite{vaidyanathan_multirate_1993}
\begin{equation}
    \label{eq:lattice-tm}
    \mbf{H}_M(z) = \mbf{R}_M \Delay(z) \mbf{R}_{M-1} \Delay(z) \cdots \Delay(z) \mbf{R}_0,
\end{equation}
where the real parameters $\theta_0, \dots, \theta_M$ define the rotation matrices $\mbf{R}_k = \mbf{R}_y(\theta_k)$ for $k=0,\dots,M$.
Fig.~\ref{fig:fir-circuit}(a) shows \eqref{eq:lattice-tm} interpreted as a single-qubit circuit. The input and output are encoded by the amplitude $\ket{0}$. The unit delay gates $\Delay(z)$ delay the amplitude $\ket{1}$ in between rotations. A projection gate $\ket{0}\bra{0}$ is added at the end to discard the second output of the filter. Fig.~\ref{fig:fir-circuit}(b) realizes the $z$-domain circuit defined in Fig.~\ref{fig:fir-circuit}(a). The buffer register has $b=\lceil \log_2(M+2) \rceil$ qubits: the buffer $k=0$ is a single-qubit work buffer, the buffers $k=1,\dots,M$ are used for unit delays, and the buffer $k=M+1$ is used to emulate the projection gate. The length of the delay used for projection is $p=2^q$, where $q$ is the size of the second qubit register. The transposition gates $\mbf{T}_k$ exchange $\ket{0}\ket{1} \leftrightarrow \ket{k}\ket{0}$.

\subsection{Parameter conversion}
\label{ssec:param-conversion}

To determine the lattice parameters, a power complementary pair of filters is required. Given an order $M$ causal FIR filter of the form $H_0(z) = a_0 + a_1 z^{-1} + \dots + a_M z^{-M}$, where $a_k \in \R$, there exists~\cite{vaidyanathan_multirate_1993} a power complement $H_1(z) = b_0 + b_1 z^{-1} + \dots + b_M z^{-M}$, where $b_k \in \R$, such that
$
    % \label{eq:power-complementary}
    |H_0(e^{i\omega})|^2 + |H_1(e^{i\omega})|^2 = 1.
$
The equation is satisfied if $H_1(z)$ is a spectral factor of $1 - |H_0(z)|^2$. The minimum-phase spectral factor can be found with homomorphic (i.e., cepstral) deconvolution~\cite{mian_fast_1982,oppenheim_discrete-time_2012}.
An implementation is available in the SciPy~\cite{virtanen_2020} library as the \texttt{minimum\_phase} function.

The parameters $\theta_0, \dots, \theta_M$ are found recursively~\cite{vaidyanathan_passive_1986}. Let
$\mbf{H}_M(z) \ket{0} = [ H_0(z), H_1(z) ]^T$
be a transfer vector of the desired response and its power complement. We write the $z$-coefficients of the transfer vector as $\ket{c_k} = [ a_k, b_k ]^T$ for $k=0,\dots,M$. The power complementary relation
%\eqref{eq:power-complementary}
implies~\cite{vaidyanathan_passive_1986} that the first and last coefficients of $\mbf{H}_M(z)\ket{0}$ are orthogonal: $ \ket{c_0} \perp \ket{c_M} $.
Therefore, an angle $\theta_M$ exists for which the $y$-axis rotation $\mbf{R}_k^{-1} = \mbf{R}_y(-\theta_M)$ sends $\ket{c_0}$ to $[\alpha, 0]^T$ and $\ket{c_M}$ to $[0, \beta]^T$, where $\alpha,\beta \in \R$ and $\alpha \ge 0$. When designing a filter with a complex transfer function, the same applies for some $2\times 2$ unitary $\mbf{R}_k^{-1}$. Since the first coefficient $\mbf{R}_k^{-1}\ket{c_0}$ becomes zero in its second row, an inverse delay $\Delay^{-1}(z)$ can be applied while keeping the system causal. This leads to the reduction $\mbf{H}_{M-1}(z) = \Delay^{-1}(z) \mbf{R}_M^{-1} \mbf{H}_M(z)$, where $\mbf{H}_{M-1}(z)$ is of order $M-1$. We proceed recursively, until $\theta_0$ is determined such that $\mbf{H}_0(z) = \mbf{R}_0$, since $\mbf{H}_0(z)$ is order zero~\cite{vaidyanathan_multirate_1993}.

\section{Simulation}
\label{sec:simulation}

The FIR filter circuit discussed in Sec.~\ref{sec:fir-filters} is implemented and tested in Qiskit~\cite{javadi_2024}. Two signals with $N=32$ samples are filtered using a lowpass FIR filter of order $M=12$ in a simulator. We demonstrate reading positive and negative values using the Hadamard test and parallel processing as outlined in Sec.~\ref{ssec:parallel-processing}. The source code for the experiment is available online\footnote{\url{https://github.com/ollpu/quantum-fir/}}.

\begin{figure}
    \centering
    \includegraphics[width=\columnwidth]{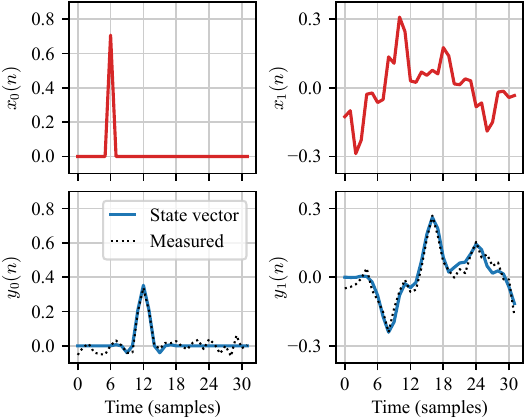}
    \caption{Two input signals (red) shown at the top, the impulse $x_0(n)$ and the randomly generated sequence $x_1(n)$, are filtered with a lowpass FIR filter ($M=12$) to produce the bottom outputs (blue), $y_0(n)$ and $y_1(n)$, respectively. The input signals are encoded in the superposition $\ket{0}\ket{A_{x_0}} + \ket{1}\ket{A_{x_1}}$ and are filtered simultaneously. Data is extracted from the simulator QSV before (red) and after filtering (blue). The dotted lines show 1024-shot measurement results. }
    \label{fig:filtering}
\end{figure}

Two input signals $x_0(n)$ and $x_1(n)$ are shown at the top of Fig.~\ref{fig:filtering}. The states are prepared as the superposition $\ket{0}\ket{A_{x_0}} + \ket{1}\ket{A_{x_1}}$ in the same circuit. Total energy is split evenly between them with a Hadamard gate. The first signal is the impulse $x_0(n) = \delta(n - 6)$, where $\delta(n)$ is the Kronecker delta. The signal is prepared as a quantum state by applying $\mbf{X}$ bit-flip gates on two qubits. The second signal $x_1(n)$ is prepared by applying $\mbf{R}_y$ gates with angles chosen uniformly at random to the $q=5$ qubits that represent the sample position $\ket{n}$ in \eqref{eq:qpam}. This generates a fractal-like signal with high-frequency content.

A windowed sinc lowpass filter is designed with a $13$-point Hamming window. The 6\,dB cutoff is set to half of the Nyquist limit, or $\pi/2$ radians per sample. The filter is not exactly bounded due to ripple, so a small $0.01\,$dB gain reduction is applied. The lattice parameters are found according to Sec.~\ref{ssec:param-conversion}. The circuit of Fig.~\ref{fig:fir-circuit}(b) is then implemented with these parameters. We choose the projection delay length $p = 32$. The qubit counts are $b=4$ and $q=5$. Since there are no measurements, it is possible to compose other kinds of processing after filtering.

Using Fig.~\ref{fig:fullsigproc}, we design a circuit to filter a signal of $N=32$ samples. The size of the circuit scales with the number of samples processed, leading to longer circuits that require fault-tolerant hardware. The test implementation is not optimal and suffers from some Qiskit limitations. Nevertheless, if the circuit is transpiled to single-qubit rotations and CNOT gates, there are $59{,}292\approx 154 NM$ gates with depth 38,387. The same circuit can process multiple signals in parallel, as shown next with two instances.

% scipy 'hamming' = 0.54 - 0.56 cos

The designed circuit is applied to the above superposition, processing the two signals in parallel into $\ket{0}\ket{A_{y_0}}+\ket{1}\ket{A_{y_1}}$ plus residual SSVs. The resulting amplitudes are shown at the bottom of Fig.~\ref{fig:filtering} in blue. The signal $y_0(n)$ contains the impulse response of the filter, while $y_1(n)$ is a lowpass version of $x_1(n)$. The output signals are delayed by 6 samples, since the filter is causal and linear-phase. The amplitudes are also estimated with the Hadamard test with 1024 shots, shown as dotted lines. QAE was not used here. Negative values are correctly observed in the second output signal $y_1(n)$.

\section{Conclusion}
\label{sec:conclusion}

A quantum circuit implementation of bounded FIR filters is presented and tested in this paper. Lossless structures, such as the introduced delay gate and rotation matrices, are identified as key QSP building blocks. The filter and the introduced state-space framework are applicable to general-purpose signal processing, and quantum speedups through parallelization and QAE are foreseeable once fault-tolerant quantum computers become available.

% \vfill\pagebreak

\section{Acknowledgments}

This work was supported by the HUCE infrastructure of the Aalto School of Electrical Engineering.
The authors are thankful to Dr.~Juha Harviainen for supervising the related master's thesis by the first author.

\bibliographystyle{IEEEbib}
\bibliography{refs}

\end{document}